\documentclass[aps,prd,reprint,groupedaddress]{revtex4-1}

\usepackage{graphicx}
\usepackage{amsmath}
\usepackage{hyperref}

\begin{document}

\title{Microwave cavity light shining through a wall optimization and experiment}

\author{Rhys G. Povey}
\email[]{rhys.povey@uwa.edu.au}

\author{John G. Hartnett}
\author{Michael E. Tobar}

\affiliation{School of Physics, University of Western Australia, WA 6009 Australia}

\date{\today}

\begin{abstract}
It has been proposed that microwave cavities can be used in a photon regeneration experiment to search for hidden sector photons. Using two isolated cavities, the presence of hidden sector photons could be inferred from a `light shining through a wall' phenomenon. The sensitivity of the experiment has strong a dependence on the geometric construction and electromagnetic mode properties of the two cavities. In this paper we perform an in depth investigation to determine the optimal setup for such an experiment. We also describe the results of our first microwave cavity experiment to search for hidden sector photons. The experiment consisted of two cylindrical copper cavities stacked axially inside a single vacuum chamber. At a hidden sector photon mass of $37.78 \; \mathrm{\mu \, e V}$ we place an upper limit on the kinetic mixing parameter $\chi = 2.9 \times 10^{-5}$. Whilst this result lies within already established limits our experiment validates the microwave cavity `light shining through a wall' concept. We also show that the experiment has great scope for improvement, potentially able to reduce the current upper limit on the mixing parameter $\chi$ by several orders of magnitude.
\end{abstract}


\maketitle

\section{Introduction}
Many extensions to the standard model of particle physics contain an extra U(1) gauge factor corresponding to a hidden sector of particles~\cite{abel, goodsell}. The only interaction between this hidden sector and normal matter is a weak kinetic mixing between the photon $\gamma$ and hidden sector photon $\gamma^\prime$~\cite{okun,holdom}. This allows for oscillations to occur between the two particles. The hidden sector photon is likely very light and belongs to a class of hypothetical particles known as WISPs (Weakly Interacting Sub-eV Particles)~\cite{jaeckel_lowenergy}. It is proposed that indirect observations of the hidden sector photon can be made by photon regeneration experiments in a similar way to ALP (Axion Like Particle) experiments~\cite{axionexp1,axionexp2,axionexp3,axionexp4}. Typically these have been laser `light shining through a wall' (LSW) experiments~\cite{BFRT1,BFRT2,BMV,ahlers_paraphoton,GammeV,ahlers_laser,afanasev_2008,fouche,afanasev_2009,ALPS,ALPS2010}. Recently however a new method for detecting hidden sector photons using microwave cavities was proposed~\cite{jaeckel_cavity}.

In microwave cavity LSW we use two resonance matched cavities separated by a wall that is impenetrable by normal photons. Electromagnetic radiation is injected into one of the cavities (the emitter cavity) and a small portion oscillates into hidden sector photons. These hidden sector photons do not interact with normal matter and are able to pass straight through the normally impervious wall. If some of these particles then oscillate back into photons inside the other (detector) cavity a signal detection could be made. The probability of this transmission taking place is~\cite{jaeckel_cavity}
\begin{eqnarray}
\mathrm{P}_\mathrm{trans} =& \frac{P_\mathrm{det}}{P_\mathrm{emit}} = \chi^4 \: Q_\mathrm{emit} \: Q_\mathrm{det} \: \left(\frac{m_{\gamma \prime} \: c^2}{\hbar \: \omega_\gamma}\right)^8 \: |\mathcal{G}|^2 \label{eq:Ptrans} \\
=& \chi^4 \: Q_\mathrm{emit} \: Q_\mathrm{det} \: \left(1 - \frac{k_{\gamma \prime}^{\: 2}}{k_\gamma^{\: 2}} \right)^4 \: |\mathcal{G}|^2 , \nonumber
\end{eqnarray}
where $P_\mathrm{det}$ and $P_\mathrm{emit}$ are the powers in and out of the respective cavities, $\chi$ is the kinetic mixing parameter, $Q$ is the cavity quality factor, $m_{\gamma \prime}$ is the hidden sector photon mass, $\omega_\gamma$ is the angular (and cavity resonance) frequency of the photons, $k_\gamma$ is the photon wavenumber, $k_{\gamma \prime}$ is the hidden sector photon wavenumber and $\mathcal{G}$ is a dimensionless function that encodes the geometric setup of the two cavities. The function $\mathcal{G}$ is a 6-integral given by~\cite{jaeckel_cavity}
\begin{multline}
\mathcal{G} \left( \frac{k_{\gamma \prime}}{k_\gamma} \right) = k_\gamma^{\:2} \int\limits_{V_\mathrm{emit}} \int\limits_{V_\mathrm{det}} \frac{\exp(i \; k_{\gamma \prime} \: |\textbf{\textsl{x}}-\textbf{\textsl{y}}|)}{4 \pi |\textbf{\textsl{x}}-\textbf{\textsl{y}}|} \\ \textbf{\textsl{A}}_\mathrm{emit} (\textbf{\textsl{y}}) \: \cdot \: \textbf{\textsl{A}}_\mathrm{det} (\textbf{\textsl{x}}) \; d^3 \textbf{\textsl{x}} \; d^3 \textbf{\textsl{y}} \label{eq:G} ,
\end{multline}
where $V$ represents the respective cavity volumes and $\textbf{\textsl{A}}$ is the normalized spatial part of the resonance electromagnetic gauge field inside the cavities satisfying
\begin{equation}
\int\limits_V \lvert \textbf{\textsl{A}} (\textbf{\textsl{x}}) \rvert^2 \: d^3 \textbf{\textsl{x}} = 1 \nonumber .
\end{equation}

The cavity Q-factors also contain geometric dependencies from the cavity geometric factor $G$,
\begin{align}
Q=\frac{G}{R_S}, & &
G = \omega_0 \frac{\int\limits_V \mu \lvert \textbf{\textsl{H}} (\textbf{\textsl{x}})\rvert^2 \; d^3 \textbf{\textsl{x}}}{\int\limits_S \lvert  \textbf{\textsl{H}}_\mathrm{T} (\textbf{\textsl{y}}) \rvert^2 \; d^2 \textbf{\textsl{y}}}\nonumber ,
\end{align}
where $R_S$ is the surface resistance, $\omega_0$ is the angular resonance frequency, $\mu$ is the permeability inside the cavity, $S$ is the surface of the cavity and $\textbf{\textsl{H}}_\mathrm{T}$ is component of $\textbf{\textsl{H}}$ tangential the cavity surface. To further study $\mathrm{P}_\mathrm{trans}$ it is convenient to define a new function that encompasses all of the geometric, electromagnetic and $k_{\gamma \prime}/k_\gamma$ dependencies. Hence we define the `full geometric function'
\begin{equation}
\mathcal{F}^2 \left( \frac{k_{\gamma \prime}}{k_\gamma} \right) = G_\mathrm{emit} \: G_\mathrm{det} \: \left(1 - \frac{k_{\gamma \prime}^{\: 2}}{k_\gamma^{\: 2}} \right)^4 \: |\mathcal{G}|^2 \label{eq:F2},
\end{equation}
measured in $\Omega^2$, such that 
\begin{equation}
\mathrm{P}_\mathrm{trans} = \chi^4 \: R_{S_\mathrm{emit}}^{\;-1} \: R_{S_\mathrm{det}}^{\;-1} \: \mathcal{F}^2\nonumber .
\end{equation}

In this paper we investigate the behavior of $\mathcal{F}^2$ for axially stacked cylinders as well as provide results of our first experimental test of microwave cavity LSW.

\section{Full geometric function}
\begin{figure}[t]
\includegraphics[width=0.3\textwidth]{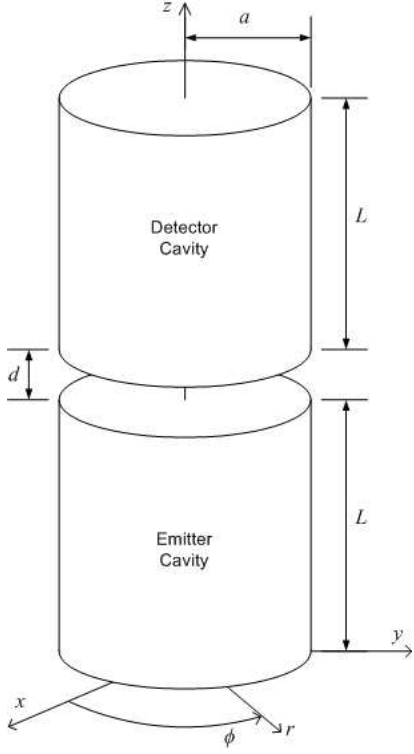}
\caption{Diagram of cavity setup with radius $a$, length $L$ and separation distance $d$.\label{fig:cylinders}}
\end{figure}

For the best chance of detection $\mathcal{F}^2$ needs to be maximized. Currently however, very little is known about the behavior of this function or its constituents. In this section we study the full geometric function from its dependence on the electromagnetic mode, aspect ratio and cavity separation. Here we only consider symmetrical cavities stacked axially as depicted in Fig.~\ref{fig:cylinders}. Both cylinders are assumed to be of the exact same dimensions. The aspect ratio ($AR$) is defined to be the diameter ($2a$) divided by the length ($L$) of each cylinder ($AR=2a/L$) and the separation distance ($d$) is defined to be the axial distance between the inside boundaries of the two cylinders. A separation distance of zero refers to the ideal and unrealizable case where the two cavities have infinitesimally thin walls and sit directly on top of each other.

Before we can calculate $\mathcal{G}$ and hence $\mathcal{F}^2$ we first need to know the electromagnetic gauge field $\textbf{\textsl{A}}$ inside the cavities. Solving Maxwell's equations inside a cylinder of radius $a$ and length (or height) $L$ we find two classes of resonance modes. The Transverse Magnetic modes with azimuthal mode number $m$, radial mode number $n$ and axial mode number $p$ are ($\mathrm{TM}_{\,m\,n\,p}$):
\begin{multline}
\textbf{\textsl{E}}_\textbf{TM} = \begin{pmatrix}
E_r \\ E_{\phi} \\ E_z \end{pmatrix} = E_0 \: e^{i \: \omega \: t} \\
\begin{pmatrix}
- \frac{p \pi}{L} \frac{a}{\varsigma_{m,n}} J_m^\prime \left( \frac{\varsigma_{m,n}}{a} r \right) \cos(m \: \phi) \sin(\frac{p \pi}{L} z)\\
\frac{m}{r} \frac{p \pi}{L} \frac{a^2}{(\varsigma_{m,n})^2} J_m \left( \frac{\varsigma_{m,n}}{a} r \right) \sin(m \: \phi) \sin(\frac{p \pi}{L} z)\\
J_m \left( \frac{\varsigma_{m,n}}{a}r \right) \cos (m \: \phi) \cos(\frac{p \pi}{L} z)
\end{pmatrix} \nonumber,
\end{multline}
\begin{multline}
\textbf{\textsl{H}}_\textbf{TM} = \begin{pmatrix}
H_r \\ H_{\phi} \\ H_z \end{pmatrix} = -i \: \omega \: \varepsilon \: a \: E_0 \: e^{i \: \omega \: t} \\
\begin{pmatrix}
\frac{m}{r} \frac{a}{(\varsigma_{m,n})^2} J_m \left( \frac{\varsigma_{m,n}}{a} r \right) \sin(m \: \phi) \cos(\frac{p \pi}{L} z)\\
\frac{1}{\varsigma_{m,n}}  J_m^\prime \left( \frac{\varsigma_{m,n}}{a} r \right) \cos(m \: \phi) \cos(\frac{p \pi}{L} z)\\
0
\end{pmatrix} \nonumber.
\end{multline} 
The Transverse Electric (TE) modes with azimuthal mode number $m$, radial mode number $n$ and axial mode number $p$ are ($\mathrm{TE}_{\,m\,n\,p}$):
\begin{multline}
\textbf{\textsl{E}}_\textbf{TE} = \begin{pmatrix}
E_r \\ E_{\phi} \\ E_z \end{pmatrix} = i \: \omega \: \mu \: a \: H_0 \: e^{i \: \omega \: t} \\
\begin{pmatrix}
\frac{m}{r} \frac{a}{(\varsigma^\prime_{m,n})^2} J_m \left( \frac{\varsigma^\prime_{m,n}}{a} r \right) \sin(m \: \phi) \sin(\frac{p \pi}{L} z)\\
-\frac{1}{\varsigma^\prime_{m,n}}  J_m^\prime \left( \frac{\varsigma^\prime_{m,n}}{a} r \right) \cos(m \: \phi) \sin(\frac{p \pi}{L} z)\\
0
\end{pmatrix} \nonumber,
\end{multline} 
\begin{multline}
\textbf{\textsl{H}}_\textbf{TE} = \begin{pmatrix}
H_r \\ H_{\phi} \\ H_z \end{pmatrix} = H_0 \: e^{i \: \omega \: t} \\
\begin{pmatrix}
\frac{p \pi}{L} \frac{a}{\varsigma^\prime_{m,n}} J_m^\prime \left( \frac{\varsigma^\prime_{m,n}}{a} r \right) \cos(m \: \phi) \cos(\frac{p \pi}{L} z)\\
- \frac{m}{r} \frac{p \pi}{L} \frac{a^2}{(\varsigma^\prime_{m,n})^2} J_m \left( \frac{\varsigma^\prime_{m,n}}{a} r \right) \sin(m \: \phi) \cos(\frac{p \pi}{L} z)\\
J_m \left( \frac{\varsigma^{\prime}_{m,n}}{a}r \right) \cos (m \: \phi) \sin(\frac{p \pi}{L} z)
\end{pmatrix} \nonumber.
\end{multline} 
Here $\varsigma_{m,n}$ (unitless) is the $n$'th root of the Bessel J function of order $m$ and  $\varsigma^\prime_{m,n}$ (unitless) is the $n$'th root of the derivative of the Bessel J function of order $m$. The parameter $\varepsilon$ is the permittivity, $\mu$ is the permeability and
\begin{eqnarray}
\omega_\mathrm{TM} &=& \sqrt{ \Bigl( \frac{\varsigma_{m,n}}{a} \Bigr)^2 + \Bigl( \frac{p \pi}{L} \Bigr)^2} \; c \: \nonumber ,\\
\omega_\mathrm{TE} &=& \sqrt{ \Bigl( \frac{\varsigma^\prime_{m,n}}{a} \Bigr)^2 + \Bigl( \frac{p \pi}{L} \Bigr)^2} \; c \: \nonumber ,
\end{eqnarray}
are the resonance angular frequencies of the cavity. $E_0$ is a constant in units of $\mathrm{V}/\mathrm{m}$ and $H_0$ is a constant in units of $\mathrm{A}/\mathrm{m}$.

Finally we find the gauge field inside the cavity satisfying both the Lorenz and Coulomb condition to be
\begin{equation}
\textbf{\textsl{A}} = \frac{i}{\omega} \textbf{\textsl{E}} \nonumber .
\end{equation}
Thus the normalized spatial part of the gauge potential $\textbf{\textsl{A}}$ appearing in Eq.~\eqref{eq:G} can be taken as the normalized spatial part of the electric field $\textbf{\textsl{E}}$ with units $\mathrm{m}^{-3/2}$.

We now have an explicit definition of $\mathcal{G}$ for any particular resonance mode we operate the pair of cavities in ($k_\gamma = \omega_\gamma / c$). Unfortunately the integral of Eq.~\eqref{eq:G} cannot be solved analytically. To understand it's behavior large numbers of numerical calculations had to be carried out. To do this, a \textsl{Mathematica}~\cite{mathematica} program was created utilizing its numerical integration features. To improve results the integrals were distributed over the component terms of $\textbf{\textsl{A}}_\mathrm{emit} \cdot \textbf{\textsl{A}}_\mathrm{det}$ and calculated separately. Furthermore the integration domains were split at the zeros of the field equations. Once evaluated these results were then used to obtain $\mathcal{F}^2$ as in Eq.~\eqref{eq:F2}.

\begin{figure}[!t]
\includegraphics[width=0.45\textwidth]{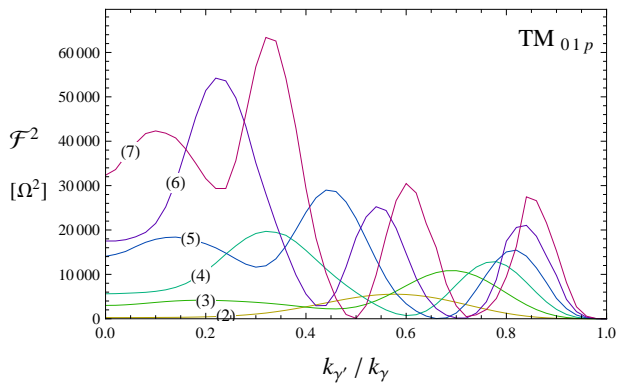}
\caption{
Curves for the $\mathrm{TM}_{\,0\,1\,p}$ mode family labeled ($p$) where $p$ corresponds to the axial mode number. Each was calculated for a cavity aspect ratio of one and zero separation distance between the cavities.
\label{fig:TM01p}}
\end{figure}

In Fig.~\ref{fig:TM01p} some typical plots of the full geometric function is given. On the scale of wavenumber ratios, $k_{\gamma \prime}/k_\gamma=0$ represents a massive hidden sector photon whose rest mass uses all of the energy of the initial photon and $k_{\gamma \prime}/k_\gamma=1$ represents a massless hidden sector photon. The $(1-k_{\gamma \prime}^{\,2} / k_\gamma)^{\,2})^4$ factor in Eq.~\eqref{eq:F2} means $\mathcal{F}^2$ will always be greatly diminished for higher wavenumber ratios.

\subsection{Mode dependency}

\begin{figure}[!t]
\includegraphics[width=0.45\textwidth]{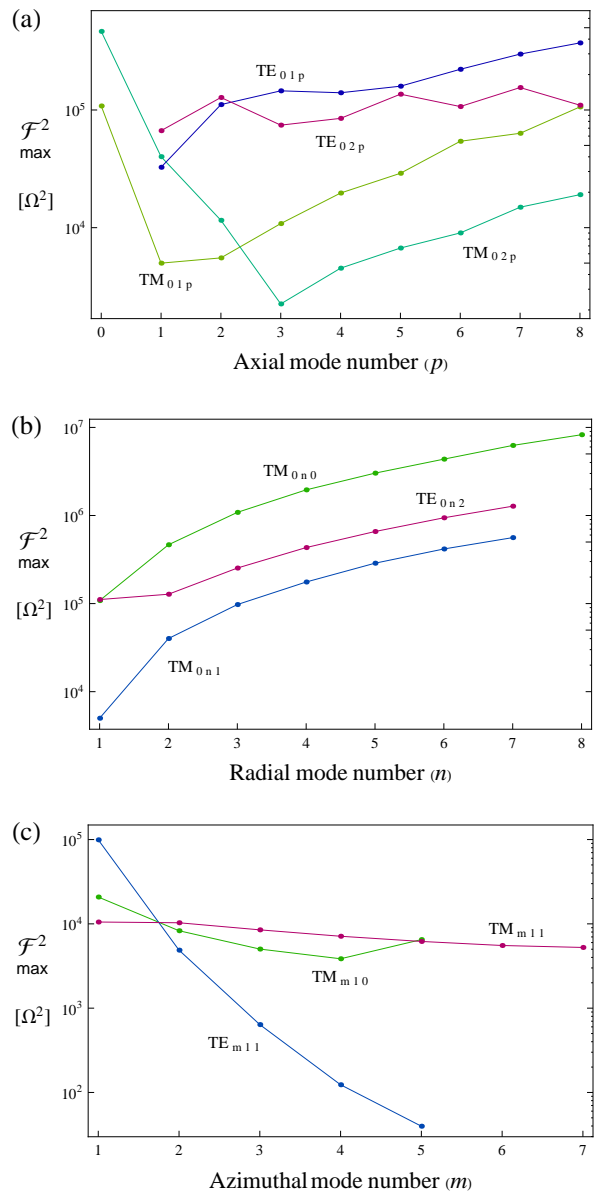}
\caption{Lines of $\mathcal{F}^2$ maximums for various families of modes against relevant mode numbers. All for a cavity aspect ratio of one and zero separation distance between the cavities. \label{fig:pnm}}
\end{figure}

The full geometric function can be very different between modes. In the investigation of mode dependencies we fix the cavity aspect ratio to one, i.e. diameter equal to length, and separation distance to zero. Figure~\ref{fig:TM01p} shows $\mathcal{F}^2$ for the family of $\mathrm{TM}_{\,0\,1\,p}$ modes and allows us to observe how $\mathcal{F}^2$ responds to varying $p$.
Taking the peak of each curve we can obtain a characteristic line for the maximum $\mathcal{F}^2$ against axial mode number. The max $\mathcal{F}^2$ line for the $\mathrm{TM}_{\,0\,1\,p}$ family of modes, as well as the $\mathrm{TM}_{\,0\,2\,p}$, $\mathrm{TE}_{\,0\,1\,p}$ and $\mathrm{TE}_{\,0\,2\,p}$ families, is shown in Fig.~\ref{fig:pnm}~(a). At higher axial mode numbers $\mathcal{F}^2$ generally tends to increase with mode number. For the TM modes however the full geometric function initially decreases to some minimum value.

The same investigation can be done for the radial mode number. The $\mathrm{TM}_{\,0\,n\,0}$, $\mathrm{TE}_{\,0\,n\,1}$ and $\mathrm{TE}_{\,0\,n\,2}$ families of modes are compared in Fig.~\ref{fig:pnm}~(b). In all cases the full geometric function increases with higher radial mode number.

Lastly, dependence on the azimuthal mode number is investigated.
In Fig.~\ref{fig:pnm}~(c) max $\mathcal{F}^2$ lines are shown for the $\mathrm{TE}_{\,m\,1\,1}$, $\mathrm{TM}_{\,m\,1\,0}$ and $\mathrm{TM}_{\,m\,1\,1}$ families of modes. The full geometric function tends to decrease with higher azimuthal mode number, most notably for the $\mathrm{TE}_{\,m\,1\,1}$ mode. For the TM modes however the change in $\mathcal{F}^2$ is much more subtle. It is uncertain whether TM whispering gallery modes with high order azimuthal mode number (often used in high-Q oscillators~\cite{locke_rsi}) in an axial stack configuration will be sensitive to hidden sector photons.

The full geometric function for each of the different modes generally occupy different regions of $k_{\gamma \prime}/k_\gamma$. This allows a range of hidden sector photon masses to be probed by different modes. We also have the option of simultaneously exciting multiple modes in the emitter cavity and covering a wider range of hidden sector photon masses at once (although at different sensitivities).

\subsection{Aspect ratio dependency}
\begin{figure}[!t]
\includegraphics[width=0.45\textwidth]{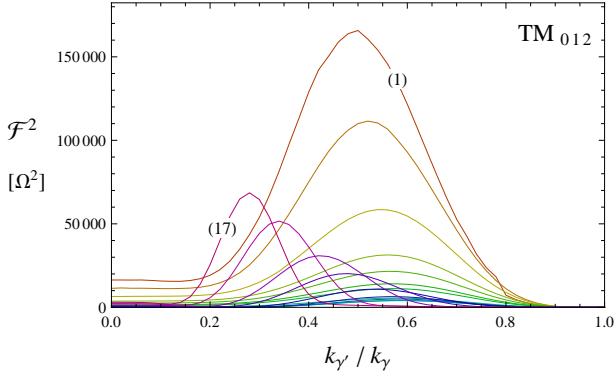}
\caption{Curves of $\mathcal{F}^2$ for the $\mathrm{TM}_{\,0\,1\,2}$ mode with aspect ratios (diameter divided by length) from (1)$1/3$ to (17)$5$, and zero separation distance between the cavities. \label{fig:ARTM012}}
\end{figure}

\begin{figure}[!t]
\includegraphics[width=0.45\textwidth]{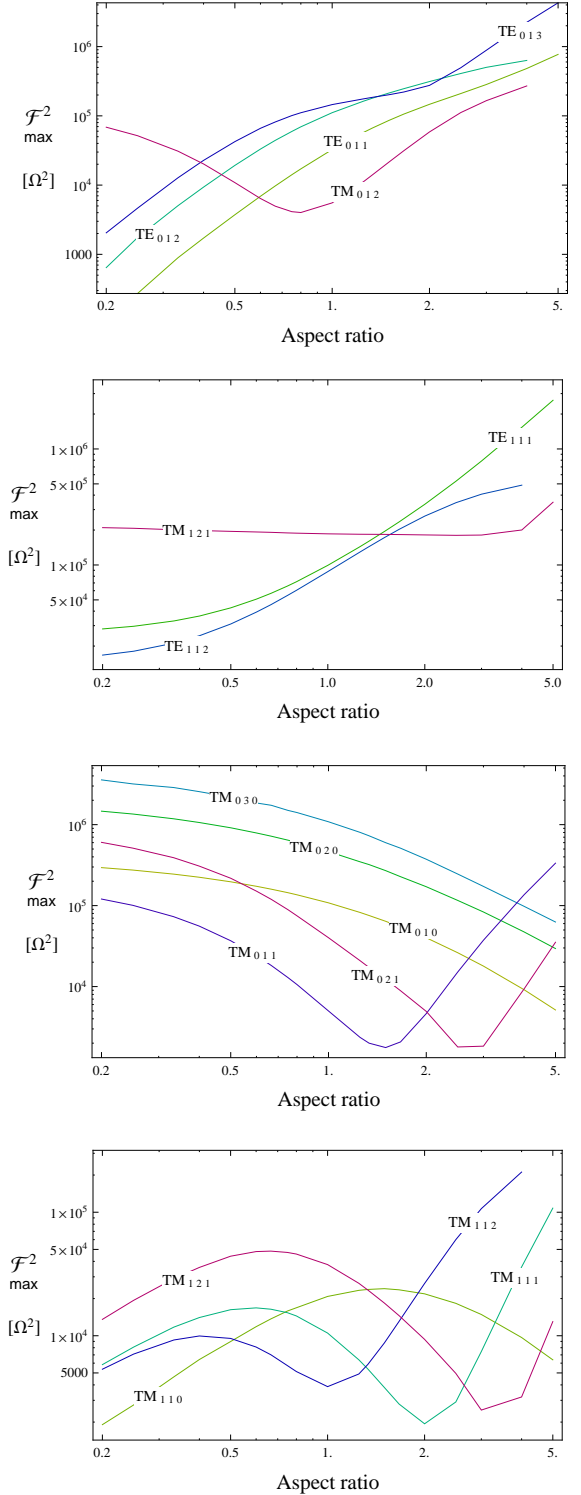}
\caption{Trends of the full geometric function maximum against aspect ratio for various modes. In all cases the separation distance between the cavities is zero. \label{fig:AR}}
\end{figure}

When the two cavities are perfectly adjacent with zero separation distance, the total size of the cavities becomes unimportant and only the aspect ratio between diameter and length affects the full geometric function. For each resonance mode the effect of varying the aspect ratio is different. Figure~\ref{fig:ARTM012} gives an example of how the full geometric function changes with different aspect ratios, in this case for the $\mathrm{TM}_{\,0\,1\,2}$ mode. In general, changing the aspect ratio not only changes the maximal value for $\mathcal{F}^2$ but also the shape and position of the peak. Nevertheless we can plot trends of the maximum $\mathcal{F}^2$ for a set of various modes as in Fig.~\ref{fig:AR}. The full geometric function seems in increase with larger aspect ratios for all modes except those with axial mode number $p=0$ which have limited $L$ dependence. Practically, extreme aspect ratio cavities are difficult to couple to and may not be usable. It is unclear if there is an optimal aspect ratio for each or some modes.

\clearpage
\subsection{Separation dependency}
\begin{figure}[!t]
\includegraphics[width=0.45\textwidth]{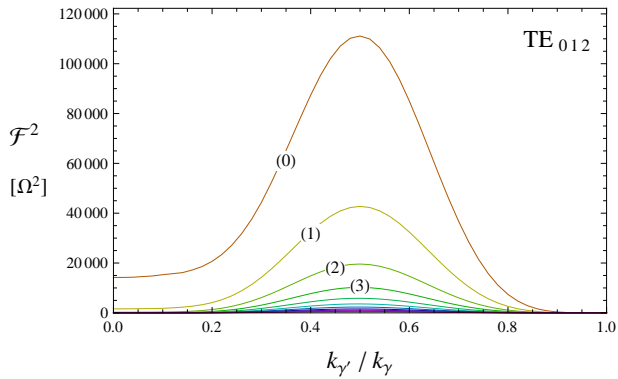}
\caption{Plots of $\mathcal{F}^2$ for the $\mathrm{TE}_{\,0\,1\,2}$ mode in cavities of size $\mathrm{length} = \mathrm{diameter} = 4 \,\mathrm{cm}$ (aspect ratio 1) with a separation distance labeled $(d)\,\mathrm{cm}$ from 0 to 10. \label{fig:dTE012}}
\end{figure}

Generally, as expected, the full geometric function decreases with greater separation distances between the two cavities. The amount by which it decreases however is different for each mode and also depends on the aspect ratio and total size of the cavities. Using the $\mathrm{TE}_{\,0\,1\,2}$ mode as an example, Fig.~\ref{fig:dTE012} shows the typical dependence of $\mathcal{F}^2$ on separation distance.

For any particular mode, both $\mathcal{G}$ and the full geometric function remain constant under proportional scaling of the cavity radius, length and separation distance,
\begin{equation}
\mathcal{F}^2_\mathrm{mode}(a,L,d)=\mathcal{F}^2_\mathrm{mode}(\alpha\:a,\alpha\:L,\alpha\:d), \nonumber
\end{equation}
where $a$, $L$ and $d$ are the cavity radius, length and separation distance respectively, and $\alpha$ is a real number greater than zero.
When $d=0$ and $a$ and $L$ are scaled together (keeping the same aspect ratio) then $\mathcal{F}^2$ remains constant as previously stated.

\begin{figure}[!t]
\includegraphics[width=0.45\textwidth]{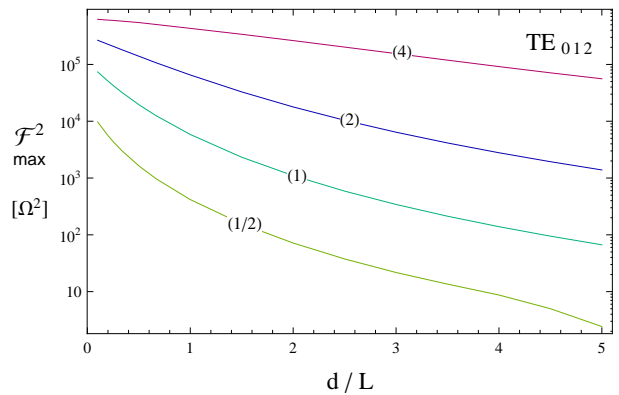}
\caption{
Trend lines of maximal $\mathcal{F}^2$ against distance over length for the $\mathrm{TE}_{\,0\,1\,2}$ mode with aspect ratios as labeled.
\label{fig:dTE012L}}
\end{figure}

In Fig.~\ref{fig:dTE012L} we plot $\mathcal{F}^2$ maximums against separation distance over cavity length ($d/L$) and compare different aspect ratios. We find that when the aspect ratio is lower (i.e. length greater than diameter) the decay in $\mathcal{F}^2$ is faster, whilst when the aspect ratio is higher (i.e. diameter greater than length) the decay in $\mathcal{F}^2$ is slower. The results are similar for other modes except when the axial mode number $p=0$ and the trend is reversed. Thus if the cavities are to be separated at large distances then a larger cavity with a greater aspect ratio is favourable.

\subsection{Optimal configuration}
From our findings in Fig.~\ref{fig:pnm} the full geometric function is optimized with the use of a $\mathrm{TM}_{\,0\,n\,0}$ mode with high radial mode number ($n$) or possibly also a $\mathrm{TE}_{\,0\,n\,p}$ mode with high axial and radial mode number. 
Figure~\ref{fig:AR} suggests that a lower aspect ratio may be better with a $\mathrm{TM}_{\,0\,n\,0}$ mode and a higher aspect ratio with a $\mathrm{TE}_{\,0\,n\,p}$ mode. 
Practical considerations will place limitations on the mode numbers and dimensions of our cavity. Firstly, we need to be able to couple effectively to the cavity and this may be difficult with obscure or extreme dimensions. Secondly, we have to consider the microwave components being used with the cavities. It is most convenient to operate in the X-band ($8-12\,\mathrm{GHz}$) range of frequencies as these are readily supported. The choice of mode and frequency will also depend on what range of hidden sector photon masses is to be explored.

For large and flat cavities the relative dependence on separation distance is the weakest. Whilst the separation should still be kept minimal, the problems of microwave leakage make it favourable to increase the separation distance to allow for better electromagnetic shielding between the cavities.

Following these guidelines it should be possible to construct an experiment exploiting a peak $\mathcal{F}^2 \sim 10^6 \; \Omega^2$ with a decent separation of $10\,\mathrm{cm}$.

\section{First experiment}

\subsection{Experimental setup}
To demonstrate the viability of microwave cavity LSW we conducted a simple experiment using two cylindrical copper cavities at room temperature. Our cavities have an internal radius of approximately $2\;\mathrm{cm}$ and internal length of approximately $4\;\mathrm{cm}$. 
The $\mathrm{TE}_\mathrm{\,0\,1\,1}$ mode was used to excite the cavities.

\begin{figure*}[!t]
\includegraphics[width=0.65\textwidth]{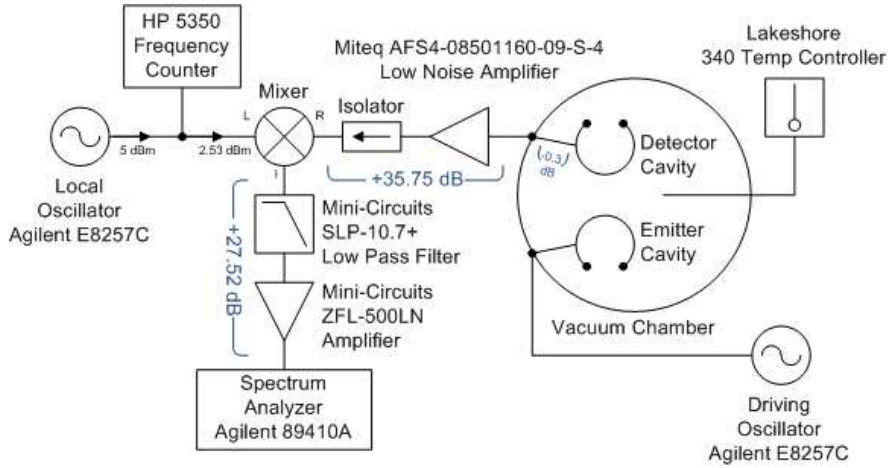}
\caption{Diagram of the microwave circuit used in our cavity experiment. \label{fig:expcircuit}}
\end{figure*}

\begin{figure}[!b]
\includegraphics[width=0.45\textwidth]{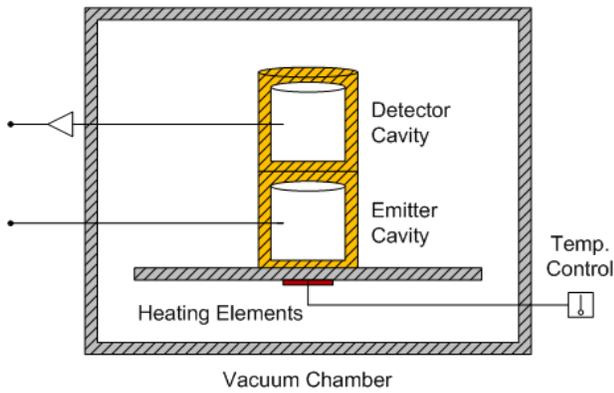}
\caption{Schematic of the experimental setup. \label{fig:expsetup}}
\end{figure}

A single loop probe was inserted in the middle of the side wall of each cavity and aligned and adjusted to maximize coupling to the $\mathrm{TE}_\mathrm{\,0\,1\,1}$ resonance mode. Operating in this mode the cavities have quality factors of $9060$ and $8370$, resonance frequencies of $9.58806\;\mathrm{G\,Hz}$ and $9.58794\;\mathrm{G\,Hz}$, resonance bandwidths of $1.01\;\mathrm{M\,Hz}$ and $1.17\;\mathrm{M\,Hz}$, and coupling coefficients of $0.97$ and $0.83$. The difference in resonance frequencies between the two cavities is $0.12\;\mathrm{M\,Hz}$, well within their resonance bandwidth of $\sim 1\;\mathrm{M\,Hz}$. The cavities were stacked axially on top of each other inside a vacuum chamber and temperature controlled to maintain the resonance frequency match. They were clamped down to provide good thermal contact. Isolation between the cavities was provided only by their individual cavity walls with no extra shielding being employed. A diagram of the cavities in the vacuum chamber is shown in Fig.~\ref{fig:expsetup}.
This setup has a peak $\mathcal{F}^2 = 9825 \; \Omega^2$ at $k_{\gamma \prime}/k_\gamma=0.3$.

To excite the emitter cavity a signal generator was used at the cavity's resonance frequency. To measure the resulting signal in the detector cavity the microwave circuit shown in Fig.~\ref{fig:expcircuit} was used. The output of the detector cavity passed through a low noise amplifier and was then mixed against a second signal generator set a few MHz off the cavity resonance frequency. This provided a signal at the offset frequency which was put through a low pass filter and preamplifier before being measured by a FFT spectrum analyzer.

\begin{figure}[!b]
\includegraphics[width=0.45\textwidth]{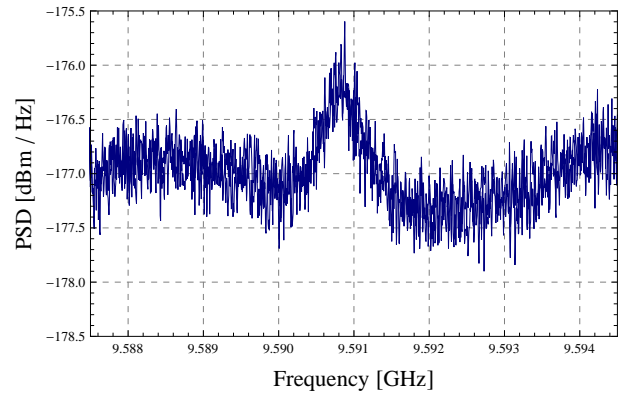}
\caption{Power spectral density showing the thermal noise of our detector cavity. \label{fig:noise}}
\end{figure}

\subsection{Limiting sensitivity}
The best possible sensitivity of our experiment will depend on the thermal noise floor of the detector cavity. The theoretical amount of Nyquist thermal noise is
\begin{equation}
N = \frac{k_B \: T}{2} \lvert \mathcal{T}(i \: \omega) \rvert^2 \nonumber,
\end{equation}
where
\begin{equation}
\mathcal{T}(i \: \omega) = \frac{2 \sqrt{\beta}}{(1+\beta)(1+2\:i\:Q\:(\omega-\omega_0)/\omega_0)} \nonumber\label{eq:transcoef}
\end{equation}
is the the transmission coefficient, in which $\beta$ is the cavity coupling coefficient, $\omega_0$ is the angular resonance frequency and $i=\sqrt{-1}$. Thus when measuring the cavity's noise spectral density ($\mathrm{Q}=8370$, $T=295\;\mathrm{K}$, $\beta=0.83$) we expect to see a Lorentzian centered at the resonance frequency with a peak value of $176.9\;\mathrm{dBm/Hz}$. Using the setup of Fig.~\ref{fig:expcircuit} the actual thermal noise measured, with an uncertainty of $\pm1.5\;\mathrm{dBm}$, is shown in Fig.~\ref{fig:noise} and agrees with our prediction.

\begin{figure*}[!t]
\includegraphics[width=0.65\textwidth]{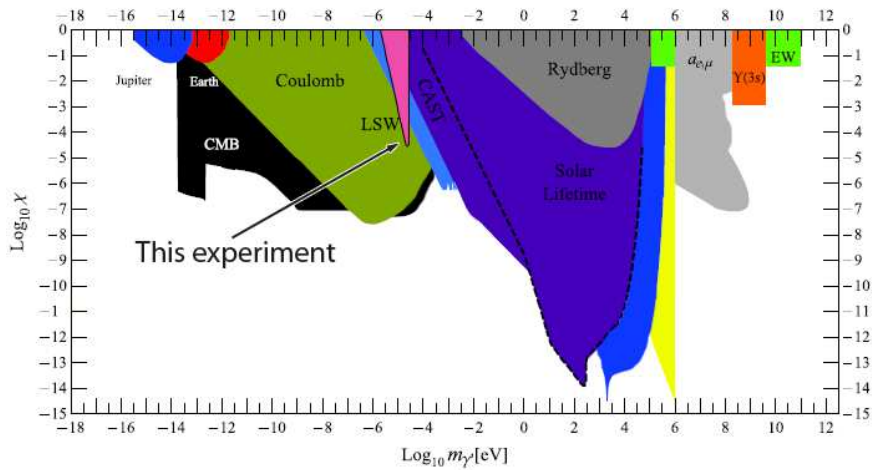}
\caption{Limits from this experiment against current hidden sector photon bounds from Ref.~\cite{goodsell}. \label{fig:expexclusion2}}
\end{figure*}

For a signal to noise ratio of one, Eq.~\eqref{eq:Ptrans} gives us a maximum sensitivity of
\begin{equation}
\chi = \left( \frac{k_B \: T}{2 \: \tau \: P_\mathrm{emit}} \right)^{\frac{1}{4}} \sqrt{\frac{R_S}{\mathcal{F}}} \label{eq:sens}
\end{equation}
where $\tau$ is the integration time and $R_S=\sqrt{R_{S_\mathrm{emit}}\:R_{S_\mathrm{det}}}$. For our experimental setup the peak $\mathcal{F}^2 / R_S^{\;2} = 1.2 \times 10^6$. If $1\,\mathrm{W}$ of incident power and an integration time of 1 week is used, then Eq.~\eqref{eq:sens} allows us to probe $\chi = 7.2 \times 10^{-9}$.

\subsection{Experimental results}
To operate the experiment various power levels from the driving signal generator ranging between $0$ and $20\;\mathrm{dBm}$ were input to the emitter cavity. As expected microwave leakage was a major problem in this simple setup. With an incident power of $0\;\mathrm{dBm}$, a reading of approximately $-66\;\mathrm{dBV}_{\mathrm{RMS}}$ was obtained from the spectrum analyzer and this scaled proportionally with higher power inputs. Taking into account the amplification and measurement system the detector cavity power output was measured to be on average $120.35 \pm 1.50 \;\mathrm{dB}$ below the power input of the emitter cavity. This still relatively large signal is most likely due to microwave leakage inside the common vacuum chamber, probably through the necessary pinhole in each cavity for vacuum pumping, unmatched SMA connections and coupling probes. 
We are unable to distinguish this signal from possible hidden sector photons and a limit can only be placed down to the strength of this signal. That is, hidden sector photons which would produce a signal greater than that measured are not observed. Using Eq.~\eqref{eq:Ptrans} we can place an upper limit on the kinetic mixing parameter $\chi$ from this experiment which peaks at $\chi=2.9\times10^{-5}$ when $m_{\gamma \prime}=3.788\times10^{-5}\;\mathrm{eV}$. A comparison of these results to previous limits on the hidden sector photon is given in Fig.~\ref{fig:expexclusion2}.


\subsection{Future work}
Our results from this prototype experiment are not an improvement on previous hidden sector photon bounds~\cite{goodsell}, but do provide promise for the future of microwave cavity LSW. Great improvements on this experiment can be made and a reduction in the $\chi$ limit by multiple orders of magnitude is possible. The two main areas for improvement are microwave leakage suppression and higher Q-factor cavities. By separating our cavities into individual vacuum chambers we can greatly reduce the amount of leakage and hence be able to place a tighter limit on the mixing parameter. This extra separation comes at the cost of reducing $\mathcal{F}^2$ but overall produces a better experiment. We have been able to determine geometries which maintain $\mathcal{F}^2 \sim 10^6 \; \Omega^2$ at separations of $10\;\mathrm{cm}$.

Higher Q-factor (lower $R_S$) emitter and detector cavities can be used to increase the probability of transmission and hence the sensitivity to $\chi$. To reap the full benefits of higher Q-factors, however, we need to be able to closely match and maintain the resonance frequency of our cavities. If two cavities with $\mathrm{Q}=10^8$ can be matched in resonance frequency at cryogenic temperatures then a fundamental sensitivity of $\chi \sim 10^{-12}$ can be achieved.

Further improvements and methods of positive signal detection from Ref.~\cite{caspers} could also be incorporated.

\begin{acknowledgments}
This work was supported by the Australian Research Council.
\end{acknowledgments}


%

\end{document}